\documentclass{aa}  

\usepackage[normalem]{ulem}
\usepackage{txfonts}
\usepackage{graphicx}
\usepackage{subfigure}
\usepackage{adjustbox}
\usepackage{hyperref}
\usepackage[switch]{lineno}
\hypersetup{
    colorlinks=true,
    linkcolor=blue,
    filecolor=blue,      
    urlcolor=blue,
    citecolor=blue,
    }
\usepackage[percent]{overpic}
\usepackage{xcolor}
\begin{document}

   \title{Suppressed "lump" EM signature in radiation pressure dominated accreting massive black hole binaries %\af{Maybe "no lump" is a bit extreme, lets' say "dimmer/suppressed"?}
   } 

   %\subtitle{I. Overviewing the $\kappa$-mechanism}

   \author{Fabiola Cocchiararo
          \inst{1,2}\fnmsep\thanks{E-mail: f.cocchiararo@campus.unimib.it} 
          Alessia Franchini\inst{2,3}\fnmsep\thanks{E-mail: alessia.franchini@unimi.it}
          Alessandro Lupi \inst{1,2,4}
          Alberto Sesana \inst{1,2}
          }

   \institute{
    Dipartimento di Fisica "G. Occhialini", Università degli Studi di Milano-Bicocca, Piazza della Scienza 3, 20126 Milano, Italy             
    \and
    INFN, Sezione di Milano-Bicocca, Piazza della Scienza 3, 20126 Milano, Italy
    \and
    Dipartimento di Fisica "A. Pontremoli", Università degli Studi di Milano, Via Giovanni Celoria 16, 20134 Milano, Italy 
    \and
    Como Lake Center for Astrophysics, DiSAT, Università degli Studi dell’Insubria, via Valleggio 11, I-22100 Como, Italy\\
       }

   \date{}

% \abstract{}{}{}{}{} 
% 5 {} token are mandatory
 
\abstract
{We investigate the impact of radiation pressure on electromagnetic signatures of accreting massive black hole binaries (MBHBs) at milli-parsec separations, using 3D hyper-Lagrangian resolution hydrodynamical simulations. We model binaries embedded in a self-gravitating circumbinary disc that evolves following an adiabatic equation of state, including viscous heating and black-body cooling. 
Focusing on binaries with a total mass of $10^6 \, M_{\odot}$, eccentricities $e=0,0.45,0.9$ and mass ratios $q=1, 0.7$, we find that radiation pressure significantly affects both the spectral energy distributions (SEDs) and the light curves (LCs). The emission from the mini-discs shifts from the optical towards UV frequencies and with a peak luminosity orders of magnitude higher, while the circumbinary disc becomes colder and dimmer as a result of its geometrically thinner configuration. Temporal variability is affected as well: near UV and soft-X ray fluxes are higher and more variable. %\af{are the fluxes higher or is the variability higher?} \fc{Both of them, we can say:"Both the flux magnitude and their temporal variability are effected as well, with higher values in the near UV and soft X-ray bands."} 
%Crucially, radiation pressure suppresses the characteristic "lump" modulation in equal-mass systems, shifting dominant periodicities to $0.36 \, f_{\rm K}$ in the equal-mass case and leaving only robust orbital period periodicities ($f=1,2 \, f_{\rm K}$) in highly eccentric binaries. 
%\af{I would say, here and in the whole paper, that a proper lump does not form in the circular case but there is still a modulation on the cavity edge timescale. In the eccentric case instead, the lump is suppressed and there is no modulation associated with the cavity orbital motion.} 
Crucially, radiation pressure suppresses the characteristic "lump" formation in equal-mass circular systems, while a lump is formed for higher eccentricities without imprinting any modulation on the flux. %regardless of eccentricity. However, 
In the circular case we still find a modulation on the cavity edge timescale at a frequency $0.36 \, f_{\rm K}$, while in eccentric binaries, only robust orbital period modulations ($f=1,2 \, f_{\rm K}$) are observed, with no modulation associated with the cavity orbital motion.
%In highly eccentric equal-mass binaries, the suppression of the lump-driven modulation enhances the detectability of its periodic signatures in surveys like the future Vera Rubin Observatory. \af{How does the suppression of the lump feature enhances the detectability?} \fc{Because 2 modulations are easier to detect and they are distinguishable at higher redshifts even taking into account a few of orbits, in contrast to the only gas pressure case. I see that this is not clear, we can say "In highly eccentric equal-mass binaries, the suppression of the lump feature removes additional modulations, enhancing the detectability of robust orbital periodicities in surveys such as the future LSST by the Vera Rubin Observatory."} \af{I do not think that the modulations are more detectable because the one related to the lump is missing.}
Moreover, the enhanced emission from the mini-discs and streams due to radiation pressure, one redshifted, results in brighter flux in the optical G band, proving detectability of MBHBs signatures even at higher redshift ($z=0.6-1.0$). Our results reveal that radiation pressure plays a crucial role in shaping MBHBs spectral and time-domain features, with implications for their identification in time-domain surveys.} 

   
\titlerunning{Radiation pressure EM signatures of MBHBs}
   \authorrunning{F. Cocchiararo et. al}
   \maketitle
%
%-------------------------------------------------------------------
%%%%%%%%%%%%%%%%% BODY OF PAPER %%%%%%%%%%%%%%%%%%
%%%%%%%%%%%%%%%%% INTRODUCTION %%%%%%%%%%%%%%%%%%
\section{Introduction}
Massive black hole binaries (MBHBs) are among the main targets of current and future gravitational wave (GW) experiments. The upcoming space-based Laser Interferometer Space Antenna \citep[LISA,][]{LISA2023} will probe the milli-hertz GW band, observing the late inspiral and merger of MBHBs with masses between $10^4-10^7 \, M_{\odot}$, across the Universe. At lower frequencies, in the nano-hertz regime, Pulsar Timing Arrays \citep[PTAs,][]{Foster1990} can detect GWs from more massive binaries at milli-parsec separations. Although GWs are essential for a definitive detection of MBHBs, identifying electromagnetic (EM) counterparts is crucial for constraining the properties of these systems, opening the era of low-frequency multi-messenger astronomy. The importance of understanding the EM emission of MBHBs has become even more compelling following the recent detections of a signal compatible with a GW background (GWB) reported by different PTA collaborations \citep[e.g.,][]{2023arXiv230616214A, nanograv2023, ppta2023,cpta2023}. While the origin of this signal can not yet be confirmed from GW data alone, identifying EM counterparts would help constraining its properties and establishing its nature \citep{Smarra2023, 2023ApJ...951L...9A, 2023ApJ...951L..11A}.

Besides the natural variability arising from the binary orbital period, one of the most promising EM signatures of MBHBs is the so-called 'lump', an over-density region orbiting at the cavity edge of the circumbinary disc with an orbital period a few times longer than the binary orbital period. Early 2D hydrodynamical simulations \citep[e.g.,][]{Macfadyen2008, dorazio2013} showed that the lump can modulate the accretion rate onto the binary, producing periodic variability in the emitted light. Subsequent 2D and 3D simulations \citep{Roedig2011, Cuadra2009, farris2014, Westernacher2022, Westernacher2023, Franchini2024a, Franchini2024b}, also including magnetic fields \citep{Krolik2010, Shi2012, Tang2018} attempted to characterise the EM signatures from MBHBs at
very small separations (i.e. tens to a hundred gravitational radii $R_{\rm g}$), confirming the lump formation and stability influence the EM emission. These results have been further validated by 3D general relativistic magneto-hydrodynamics (GRMHD) simulations \citep{Krolik2010, Noble2012,Bowen2018}. 

Despite the diversity in numerical methods and physical models adopted, all these studies consistently find that the lump originates when the particles of gas flung back to the cavity edge, shock and deposit material at the inner edge of the circumbinary disc. Moreover, the lump is stronger in circular equal-mass binaries, with amplitude decreasing for lower mass ratios \cite{Westernacher2022}. However, at larger binary separations (e.g., $10^{-4}\, \rm pc \sim 10^4 R_{\rm g}$) \cite{Cocchiararo2024} highlighted that the lump can also form in eccentric equal-mass binaries, thus making it a potential smoking-gun signature of equal-mass systems.

It is therefore important to verify whether this dynamical feature survives the implementation of additional physics that is expected to play an important role in the problem. One additional %such 
piece of physics is radiation pressure, whose impact at large  binary separations has remained largely unexplored.
%However, the impact of additional physical processes such as radiation pressure at higher binary separation has remained largely unexplored. 
In %\al{I would remove "our recent work"} 
\cite{Cocchiararo2025}, we addressed this gap by investigating its impact on the evolution of accreting MBHBs at milli-parsec scales using 3D numerical simulations.
We found that radiation pressure alters the vertical and thermal structure of the disc, producing a geometrically thinner, and therefore colder, configuration. This leads to a reduced accretion rate onto the binary and suppresses cavity eccentricity growth and precession in circular equal mass binaries, with potential implications for the EM emission and periodic signatures used to identify these systems. Consistently, 3D radiation magneto-hydrodynamics (RMHD) simulations of MBHBs at smaller radii reported similar effects, further supporting the key role of radiation pressure in shaping circumbinary disc dynamics \citep{Tiwari2025}.

A detailed characterisation of EM signatures at intermediate separations of  $10^{-4}-10^{-2}\, \rm pc$, corresponding to orbital periods of days to years, 
in particular under the influence of radiation pressure, is currently missing and it is essential both to constrain the origin of the GWB evidence in PTA experiments and to identify MBHB candidates in time-domain surveys, such as the future upcoming Vera Rubin Observatory  \citep[VRO;][]{LSST2009}. These candidates may represent the progenitors of merging binaries that will be detected by LISA
\citep{Charisi2016, xin2024, xin2025}.

Motivated by these considerations, in this work we have computed for the first time the spectral energy distributions (SEDs) and the multi-wavelength light curves (LCs) from 3D numerical simulations with hyper-Lagrangian refinement of milli-parsec scale binaries including the radiation pressure contribution in the disc. % in our model. 
We use the same set of numerical simulations presented in \cite{Cocchiararo2025}. The simulations include %the contribution account for the thermodynamics evolution of the gas using 
a radiative cooling prescription in the form of black body cooling. They also include gas self-gravity %a self-gravitating circumbinary disc 
and the Shakura-Sunyaev prescription for viscosity \citep{ShakuraSunyaev1973}. 
We computed the electromagnetic emission for three values of the binary eccentricity $e=0,0.45,0.9$ and mass ratio $q=1,0.7$. The simulated binaries were placed at different redshifts ($z=0.1,0.4,0.7,1.0,1.5$), and the resulting flux was evaluated across different frequency bands, with particular emphasis on the optical bands that will be accessible to the upcoming VRO. 
Since we explore the same set of parameters used in the set of simulations presented in \cite{Cocchiararo2024}, in which radiation pressure was not included, we can directly estimate its effect on the electromagnetic emission.
%We compare our results with what found by simulations that include only gas pressure presented in \citep{Cocchiararo2024}. 

The paper is organised as follows: in Sec. \ref{sec:NumModel}, we briefly describe the numerical details of the simulations and the physical parameters we use, the circumbinary disc physics assumptions, and their thermal emission calculation. In Sec.\ref{sec:results}, we show and discuss the main results we obtained and in Sec. \ref{sec:conclusions} we draw our conclusions and discuss possible observational implications. 

%%%%%%%%%%%%%%%%%%%%%%%%%%%%%%%%%%%%%%%%%%%%%%%%%%%%%%%%%%%%%%%%%%%%%%%%%%%%%%%%%%%%%%
%%%%%%%%%%%%%%%%%%%%%%%% PHYSICAL AND NUMERICAL SETUP %%%%%%%%%%%%%%%%%%

\section{Numerical and physical set up} 
\label{sec:NumPhyModel}
\subsection{Binary and circumbinary disc model}
\label{sec:NumModel}

We use the same set of simulations presented in \cite{Cocchiararo2025}, performed with the 3D meshless finite mass (MFM) version of the code {\sc gizmo} \citep{Hopkins2015}. The simulations include the adaptive particle splitting to increase the resolution inside the cavity carved by the binary, following the technique described in \cite{Franchini2022}. The refinement scheme is the same used in our previous work \cite{Cocchiararo2024}. 

The total mass of the live binary \citep{franchini2023} is $M_{\rm B} = M_{\rm 1} + M_{\rm 2} = 10^6 \, M_{\odot}$ where $M_{1}$ and $M_{2}$ are the masses of the primary and secondary black hole respectively, and the initial separation is $a_{\rm 0}=4.8\times 10^{-4} \, \rm{pc} \simeq 1.2 \times 10^4 \, R_{\rm g}$, where $R_{\rm g} = GM_{\rm B}/c^2$ is the gravitational radius of the binary. The binary components are modelled as sink particles with accretion radius $R_{\rm sink}=0.05 a_0 = 2.4 \times 10^{-5} \, \rm pc$. We initially modelled the circumbinary disc with $N= 10^6$ gas particles for a total mass of $M_{\rm D} = 0.01 M_{\rm B}$, characterised by a surface density profile $\Sigma \propto R^{-1}$ between between $R_{\rm in}=2a$ and $R_{\rm out}=10a$ in the circular cases and between $R_{\rm in}=3a$ and $R_{\rm out}=10a$ in the eccentric cases.
The disc is coplanar with the binary orbit and has an initial aspect ratio $H/R = 0.1$. 

The gas evolves following an adiabatic equation of state with $\gamma = 5/3$. Angular momentum transport within the disc is modelled through a Shakura-Sunyaev $\alpha$-viscosity prescription \citep{ShakuraSunyaev1973}, with $\alpha=0.1$. We also include the gas self-gravity \citep{lodato2007sg} and we initialise the disc in a gravitationally stable state by setting the initial Toomre parameter $Q>1$ \citep{Toomre1964}. Finally, simulations include black-body-like radiative cooling and the contribution of radiation pressure.

We performed six simulations, three of which accounted for radiation pressure. We explore different combinations of binary eccentricity and mass ratio: $e=0$ and $q=1$, $e=0.45$ and $q=0.7$, $e=0.9$ and $q=1$. The equal-mass binaries are evolved for 1300 binary orbits, while the unequal-mass cases were extended to 1600 binary orbits, ensuring that the disc reaches a quasi-steady state, defined as the stage when the ratio between the change in binary angular momentum over accretion rate no longer varies with time.

In order to better capture the emission properties of the mini-discs and streams, we re-ran the last $400$ binary orbits of each simulation with radiation pressure at higher resolution within the cavity. We compared these results %both lower and higher 
with those obtained at our fiducial resolution and we found no substantial differences in the main features, either in the dynamical evolution of the binary and disc or in the flux periodicities. 

For a full description of the numerical and physical setup, we refer the reader to \cite{Cocchiararo2025}. 
%%%%%%%%%%%%%%%%%%%%%%%%%%%%%% Emission model %%%%%%%%%%%%%%%%%%%%%%%%%%%%%%%%%%%%%%%%%%%%%
\subsection{Emission model}
\label{sec:EmissionModel}

We compute the disc temperature, emission and resulting light curves following the same method used in our previous work \citep{Cocchiararo2024}.
For each gas particle $i$, the temperature is calculated by numerically solving the total pressure equation accounting for both the gas and radiation pressure contributes to the total pressure: 

\begin{equation}
    \label{eqn:Ptot}
    P_{\rm tot} = P_{\rm gas} + P_{\rm rad} = \frac{\rho k_{\rm B} T_{\rm i}}{m_{\rm p}\mu} + \frac{4}{3}\frac{\sigma_{\rm SB} T_{\rm i}^4}{c}.
\end{equation}

We divide the disc temperature domain into a 2D matrix with dimensions of $512 \times 512$ pixels in the $x$-$y$ plane, which coincides with the MBHB orbital plane. For each pixel, we compute the midplane temperature as the average temperature of all the particles within the $z$ co-ordinate $-0.05 a < z < 0.05 a $, obtaining the midplane temperature matrix, $T_{\rm c}$. 
Assuming that the disc is optically thick ($\kappa\Sigma > 1 $, we do not model optically thin regions: we assume optical thickness larger than 1 and neglect any effect related to a possible atmosphere around the disc), we then computed the effective temperature as:
\begin{equation}
    \label{eqn:Teff}
    T_{\rm eff}^4 = \frac{4}{3}\frac{T_{\rm c}^4}{\kappa \Sigma}
,\end{equation}
where $\Sigma$ is the 3D surface density\footnote{In our model, $\Sigma$ is computed in 3D using the Sobolev length approximation as $\Sigma \sim \rho^2 / |\nabla{\rho}| + \rho*h/N_{\rm ngb}$ where $\rho$ is the 3D volumetric density, $|\nabla(\rho)|$ is the norm of the gradient, $h$ is the smoothing length, and $N_{\rm ngb}$ is the number of neighbours.} of each element of the matrix. 

%\af{Add here the details that Krolik was mentioning about the column density being 2D, etc.}

We compute the luminosity emitted by each pixel using Planck's law for black-body radiation:
\begin{equation}
    \label{eqn:planck}
    dL_{\nu} \equiv B_{\nu }  \, {\rm d}\nu {\rm d}A  = \frac {2h \nu^3}{c^2}\, \frac{{\rm d}\nu}{  \exp\left({\frac{h\nu}{k_{\rm B}T_{\rm eff}}}\right) -1 } {\rm d}A 
,\end{equation}
where $h$ is the Planck constant, $k_{\rm B}$ is the Boltzmann constant, and d$A$ is the area of each element.
To isolate the emission from each part of the disc, we divide the simulated spatial domain into five different regions: the two mini-discs that extend from the sink radius of each component to the Roche Lobe size, the streams region that extends from outside the Roche Lobe out to $r = 3a$, an inner part of the disc, $3a < r < 5a $, and an outer part of the disc,  $5a < r < 10a $. 
For each region, we compute the SEDs through the sum of each pixel luminosity obtained with Eq.~\eqref{eqn:planck}. 

We also include the non-thermal emission expected from the hot corona, using the empirical correction to the bolometric luminosity based on observations in \cite{Duras2020} in the hard X-ray (i.e. $0.2 - 10 \, \rm{keV}$) band. 
Due to the lack of a detailed model for the non-thermal corona in the accretion discs that form around each binary component, we assume the mini-discs to behave similarly to single MBH AGN discs, where the X-ray emission is produced by the corona.
We assume the corona emission follows a power law spectrum with $\nu L_\nu\propto\nu^{-0.7}$ \citep{Regan2019} and its emission $L_{\rm X}$ is proportional to the total luminosity of the mini-discs $L_{\rm MDs}$ through the correction factor $K_{\rm X}$, computed as: 
\begin{equation}
    \label{eqn:kx}
    K_{\rm X}(L_{\rm MDs}) = a \left [ 1 + \left (\frac{\log(L_{\rm MDs}/L_{\odot})}{b} \right )^c \right ]
,\end{equation}
where $a,b,c$ are best-fit parameters shown in Table 1 in \citep{Duras2020}. 

Finally, by placing the sources at different distances from the observer and assuming isotropic emission, we compute the total flux in different bands integrating the redshifted observed flux, $f_{\nu}(\nu_{o})$, over a specific range of frequencies:

\begin{equation}
    \label{eqn:flux}
    F = \int_{\nu_1}^{\rm \nu_2} f_{\nu}(\nu_{o}) \, d \nu_{\rm o} = \int_{\nu_1}^{\rm \nu_2} \frac{L_{\rm \nu}(\nu_e)}{4\pi d_{\rm L }^2}(1+z) \, d \nu_{\rm o},
\end{equation}
where $L_{\nu}(\nu_{e})$ is the luminosity as a function of the frequency, $\nu_e$, and $d_{\rm L }$ is the luminosity distance \citep{Hogg2000, hogg2002k}.%\af{is this $l_\nu$ the $L_\nu $ in eq. 3? if yes, uniform the notation.}

To evaluate if the variability of the flux in the optical bands is observable, we add a white Gaussian noise with standard deviation corresponding to the Vera Rubin Observatory $5\sigma$ sensitivity threshold in the considered band (see \url{https://pstn-054.lsst.io/PSTN-054.pdf} for details). We then apply a fast Fourier transform (FFT) over a limited number of cycles, %consistent with 
assuming a survey duration of approximately 10 years, to quantitatively evaluate the presence and origin of any spectral peaks. A more detailed statistical analysis of the detectability of these features will be presented in F.Cocchiararo et al (in prep).

%%%%%%%%%%%%%%%%%%%%%%%% % RESULTS %%%%%%%%%%%%%%%%%%%%%%%%%%%%%%%%%%%%%%%%%%%%%%% 
\section{Results}
\label{sec:results}
We here present the results we obtained from our numerical simulations, performed both neglecting and including the radiation pressure contribution, as described in Section \ref{sec:NumPhyModel}. In particular, we focus on the disc emission and on the LCs in various frequency bands, and we investigate how radiation pressure impacts the EM observational signatures with different binary configurations. 
We note that the plots showing the flux in different frequency bands and their FFT analysis for gas-pressure-only simulations with equal mass binaries have already been presented in Figure 4 and Figure 5 of \cite{Cocchiararo2024}. In this work, we instead show the results obtained when including radiation pressure contribution for the following binary configurations: $e=0$ and $q=1$, $e=0.45$ and $q=0.7$, and $e=0.9$ and $q=1$. The case with $e=0.45$ and $q=0.7$ and only gas pressure is briefly discussed for comparison purposes. 

\subsection{Spectral energy distributions}
\label{sec:SEDs}

\begin{figure}
    \begin{center}
    \includegraphics[width=\columnwidth]{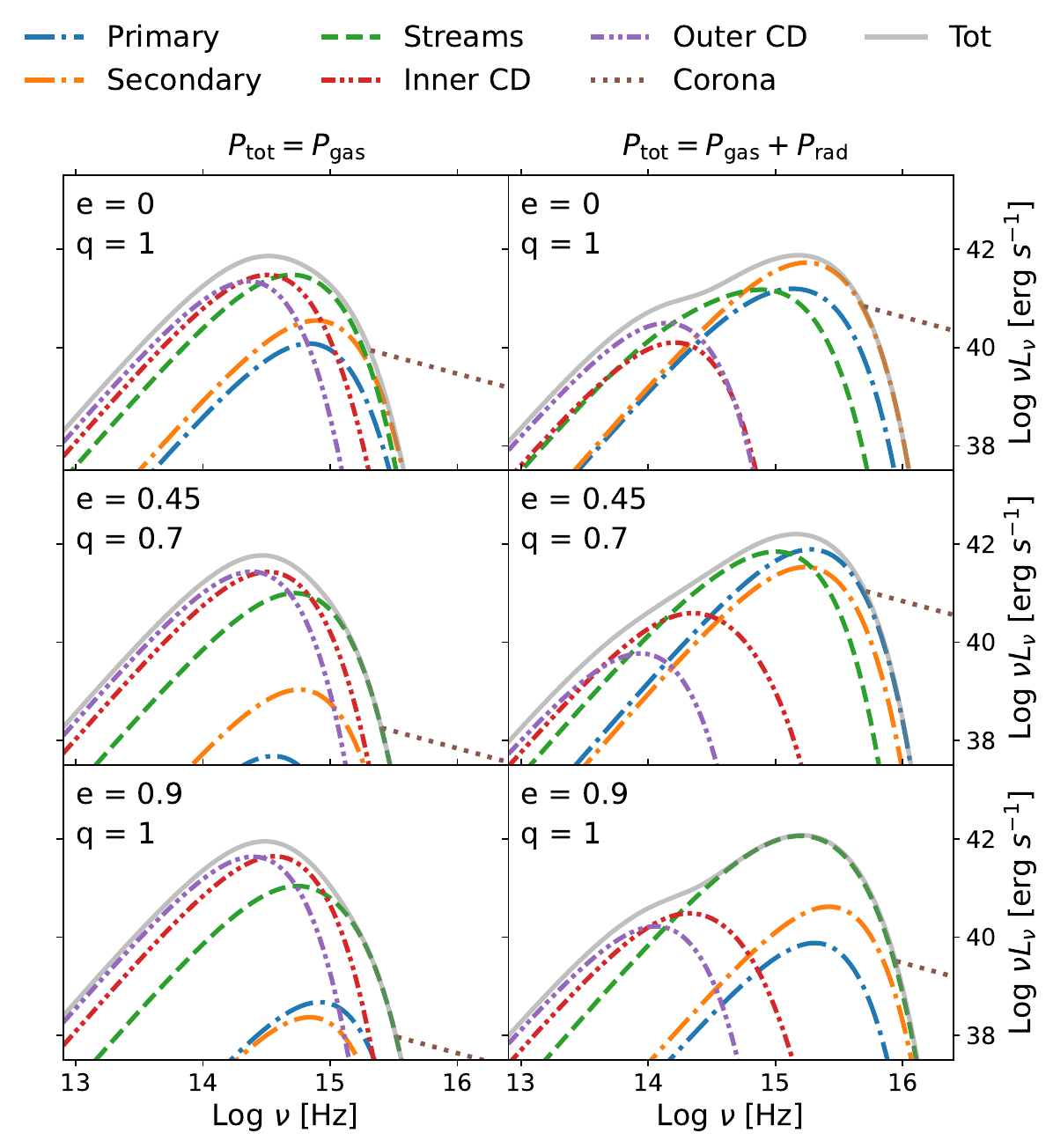}
    \caption{Spectral energy distributions (SEDs) obtained by including the contribution of the radiation pressure either a posteriori (left column) or during the binary evolution (right panel) at the same accretion rate for  the $e=0$ $q=1$ binaries (top panel), $e=0.45$ $q=0.7$ binaries (centre panel), and $e=0.9$ $q=1$ binaries (bottom panel). The contribution from the different regions of the disc (e.g. mini-discs, stream, inner and outer part of the circumbinary disc and the corona) are highlighted in different colours and have been computed at the time  when the accretion rate of both binaries is equal. %\af{I would remove the "(+Prad)" label. We can just say in the text that the spectrum in the Pgas only case is calculated assuming that the contribution of Prad is not negligible.}
    }
    \label{fig:SEDs}
    \end{center}
\end{figure}
%%%%%%%%%%%%%%%%%%%%%%%%%  PROFILES %%%%%%%%%%%%%%%%%%%%%%%%%%%%%%%%%%%%%%%%%%%%%%%%%%
\begin{figure}
    \begin{center}
    \includegraphics[width=\columnwidth]{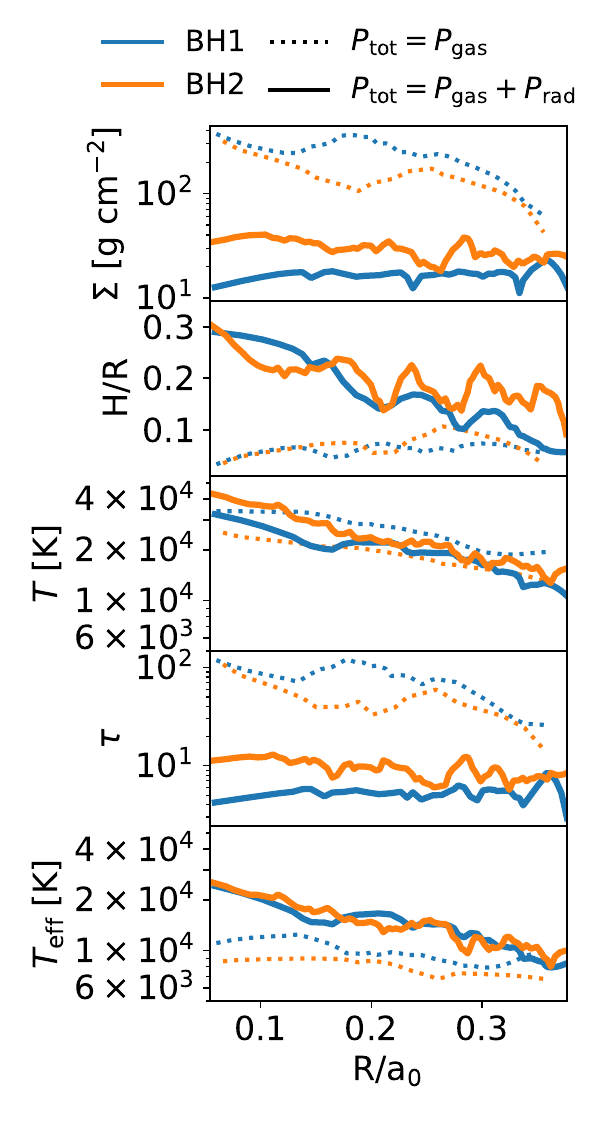}
    \caption{Profiles of surface density (first row), disc aspect ratio H/R (second row), midplane temperature (third row), optical depth $\tau$ and the effective temperature (last row) for the two mini-discs around each black hole in the circular equal mass binary, at time $t=1133 \, P_{\rm B}$, the same shown in Fig. \ref{fig:SEDs}. Blue and orange lines correspond to the two mini-discs while the solid and dashed lines refers to the simulation without and with the implementation of the radiation pressure, respectively. We calculate the profiles of these quantities within the Roche Lobe of each black hole.
    We computed the midplane temperature and the effective temperature assuming that both the gas and the radiation pressure contribute to the hydrostatic equilibrium of the disc.}
    \label{fig:profiles}
    \end{center}
\end{figure}
%%%%%%%%%%%%%%%%%%%%%%%%%%%%%%%%%%%%%%%%%%%%%%%%%%%%%%%%%%%%%%%%%%%%%%%%%%%

%%%%%%%%%%%%%%%%%%%%%%%%%%%%%Figure LCs %%%%%%%%%%%%%%%%%%%%%%%%%%%%%%%%%%%%% 

\begin{figure*}
    \begin{center}
    \includegraphics[width=2\columnwidth]{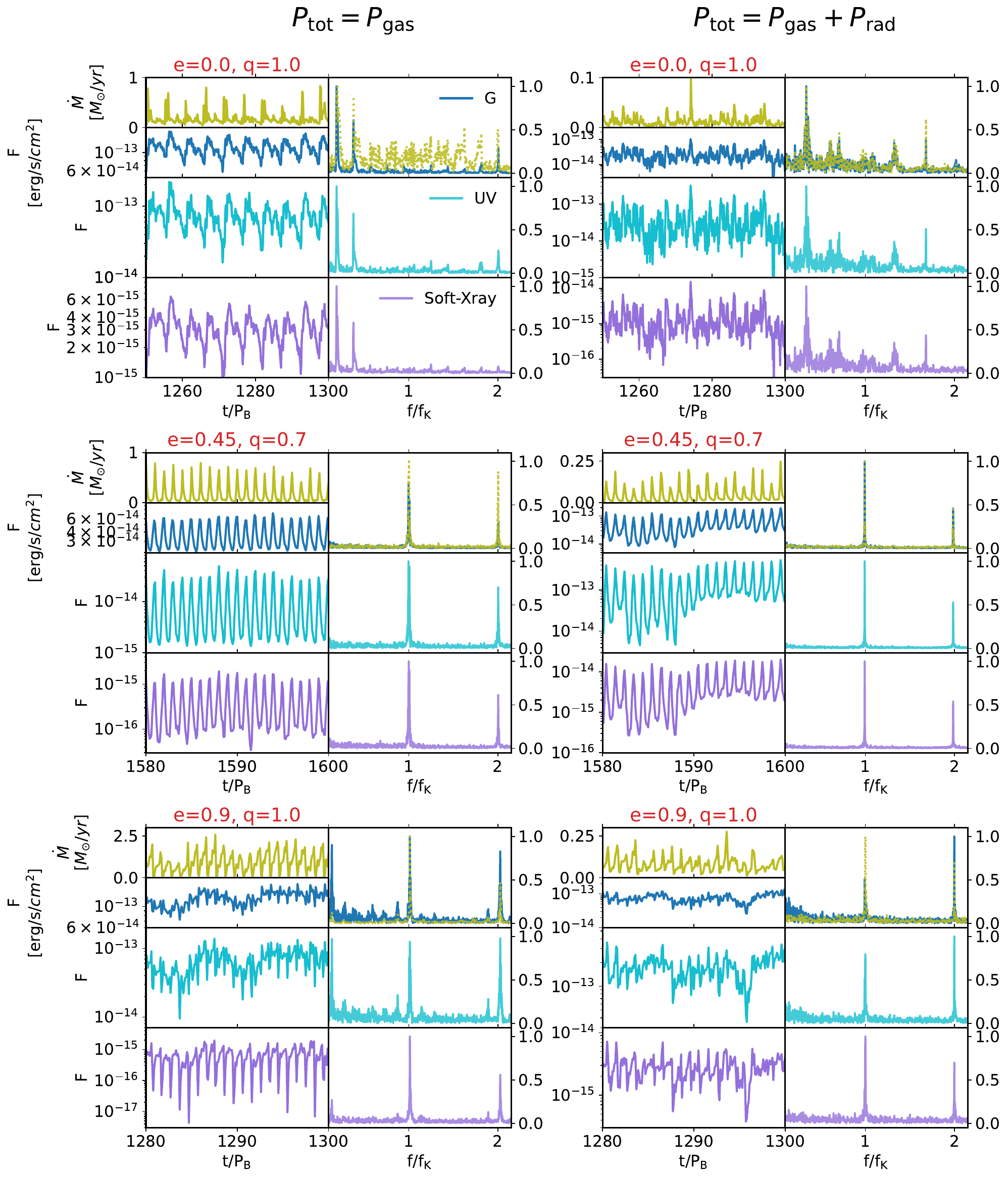}

    \caption{Light curves and the corresponding Fast Fourier Transform (FFT) obtained by including the contribution of the radiation pressure either a posteriori (left column) or during the binary evolution (right panel) for the $e=0$ $q=1$ binaries (top panel), $e=0.45$ $q=0.7$ binaries (centre panel), and $e=0.9$ $q=1$ binaries (bottom panel). In each panel, the first row shows the accretion rate (green line) and the optical G flux (blue line), while the second and last row show the flux and FFT in the UV (light blue line) and soft-X (purple line) band, respectively. The FFT is computed over 300 orbits in the window $t=1000-1300 \, P_{\rm B}$ for the circular case and over 400 orbits in the window  $t=1200-1600 \, P_{\rm B}$ and $t=900-1200 \, P_{\rm B}$ for the mildly and highly eccentric cases, respectively. The FFT is normalised to unity and shown as a function of $f/f_{\rm k}$ with $f_{\rm k}$ the Keplerian frequency of the binary. The optical flux is computed considering an extra Gaussian noise component, as described in Sec. \ref{sec:EmissionModel}.}
    \label{fig:LCs_FFT}
    \end{center}
    
\end{figure*}
%%%%%%%%%%%%%%%%%%%%%%%%%%%%%%%%%%%%%%%%%%%%%%%%%%%%%%%%%%%%%%%%%%%%%%%%%%% 

We compute the EM emission from our numerical simulations using Plank's law (Eq. \ref{eqn:planck}). In determining the disc temperature, we account for both the gas and radiation pressure contribution, as explained in Section \ref{sec:EmissionModel}. We divide our spatial simulated domain into five different regions: the two mini-discs extending from the sink radius of each component to the Roche Lobe size, the stream region that extends from outside the Roche Lobe to $r = 3a$, an inner part of the disc, $3a < r < 5a $, and an outer part of the disc,  $5a < r < 10a $. For each of these regions, we obtain the total SED  by integrating the flux emitted by each pixel over the entire spatial domain. 
Additionally, we add a posteriori a non-thermal contribution from a hot corona with a luminosity proportional to that of the mini-discs that form around each individual black hole (see Sec. \ref{sec:EmissionModel} for method details).

Figure \ref{fig:SEDs} shows the SEDs obtained for each region for equal mass binaries $e=0$ and $q=1$ (top panel), unequal mildly eccentric binaries $e=0.45$ and $q=0.7$ (middle panel) and equal highly eccentric binaries $e=0.9$ and $q=1$ (bottom panel), from simulations without and with radiation pressure, on the left and right column respectively. The spectra are shown at a the same time-step.
The spectra obtained by analysing the emission from the mini-discs and the streams region (blue, orange, and green lines) exhibit a clear shift of the emission peak toward higher frequencies when radiation pressure is included, transitioning from peak emission in the optical to the UV band. In the equal-mass circular case ($e=0$, $q=1$), the minidiscs show an increase in peak luminosity of approximately an order of magnitude, %$\sim 1$ unit in $\log L_{\nu}$,
with peak frequency shifting by $\log\, (\nu/\rm Hz) \sim 0.3$. The stream region, instead, shows a frequency shift $\log\, (\nu/\rm Hz) \sim 0.2$ but does not show a significant change in luminosity. 
For the eccentric cases the effect of radiation pressure is even more pronounced. In the moderately eccentric case ($e=0.45$, $q=0.7$), both primary and secondary minidiscs display a significant increase in peak luminosity by $\sim 3$ orders of magnitude, with a frequency shift of the emission peak of $\log\, (\nu/\rm Hz) \sim 0.6$ and $\log\, (\nu/\rm Hz) \sim 0.4$, respectively. The stream region also shifts toward UV frequencies ($\Delta\log\, (\nu/\rm Hz) \sim 0.3$) and increases by a factor of $\sim 1$ in luminosity. 
In the highly eccentric case ($e=0.9$, $q=1$), the mini-discs still show an enhancement in peak luminosity by $\sim 2$ orders of magnitude, with a shift in frequency of $\log\, (\nu/\rm Hz) \sim 0.4$ and $\log\, (\nu/\rm Hz) \sim 0.6$ for the mini-disc around the primary and secondary, respectively. 

%In contrast, 
We find that the inner and outer regions of the circumbinary disc (red and purple lines) follow an opposite trend. In all cases, the inclusion of radiation pressure causes their peak emission to shift towards lower frequencies by approximately $\log\,(\nu/\rm{Hz}) \sim 0.2-0.3$, while the luminosity decreases by $\sim 1$ order of magnitude. This behaviour is consistent with the fact that radiation pressure significantly alters the vertical and thermal structure of the circumbinary disc, resulting in a geometrically thinner, therefore colder configuration \citep{sadowski2011, mishra2016, Cocchiararo2025, Tiwari2025}. In \cite{Cocchiararo2025}, we found that the disc aspect ratio settled around $H/R \sim 0.02$ and the effective temperature dropped to $T_{\rm eff} \sim 2-3 \times 10^3 \, \rm K$, a factor $\sim 3$ lower than in gas-only simulations. 
 
Figure \ref{fig:profiles} shows the radial profiles of surface density $\Sigma$, disc aspect ratio $H/R$, midplane temperature $T$, optical depth $\tau$ and effective temperature $T_{\rm eff}$ for the two mini-discs around each binary component for the circular equal mass case, at time $t=1133 \, P_{\rm B}$, for the simulation with and without radiation pressure (straight and dotted lines respectively). 
The inclusion of radiation pressure in the simulation causes the surface density to decrease by 1 order of magnitude compared to the gas-only case, while $H/R$ increases by a factor  $\sim 3$. The mini-discs experience a vertical expansion because of their lower density. % and the surface density are lower \af{with }respect to the ones in the \af{gas-only} run and the circumbinary disc in the radiation pressure run \citep[for details, see Fig. 2 in][]{Cocchiararo2025} and 
Since the radiation pressure provides most of the hydrostatic support, a decrease in density causes an increase in aspect ratio, if the temperature is roughly the same.  
The midplane temperature remains indeed comparable in both cases ($T\sim 2-4\times 10^4 \, \rm K$).
%but, the high dominance of radiation pressure respect to the gas pressure in the mini-discs ensures a 
The decrease in surface density leads to a lower optical depth ($\tau$ decreases from $\sim 10^2$ to $\sim 100$), which translates into a higher effective temperature: $T_{\rm eff}$ rises from $\sim 6 \times 10^3-10^4 \rm K$ in the gas-only simulation to $\sim 2 \times 10^4-3\times10^4 \rm K$ when radiation pressure is included.
A lower surface density would also imply a more efficient black-body cooling. However the mini-discs continuously experience shocks from the streams and during their mutual interactions, resulting in heating overcoming radiative cooling.

\subsection{Light curves}
\label{sec:LCs}

We compute the integrated luminosity from each region of the disc to obtain the light curves (LCs), which trace the time evolution of the flux. For simplicity, we present the LCs for a source at redshift $z=0.1$ and then we explore how redshift affects the detectability of a source. 
The bolometric flux is calculated in three different regions of the EM spectrum: in the optical, using the VRO filters frequency bands (see Table 1 in \cite{Cocchiararo2024} for frequency band value specifications), in the near-UV ($1.0-1.5 \times 10^{15}$ Hz or $4.13-6.20$ eV), and in the soft X-ray ($7.25-48.3 \times 10^{16}$ Hz or $0.3-2$ keV).
We show the results in the optical G filter only because, as shown in our previous work \cite{Cocchiararo2024}, the amplitude of the peak in the FFT of the light curve varies across the different optical bands, reaching its maximum in the optical G band, making it the most suited for identifying periodic modulations from MBHBs in VRO. This choice is further supported by the results presented in \cite{Chiesa2025}, which indicate that the VRO G band is expected to yield the highest number of detectable MBHB candidates. 

The results for simulations without and with the implementation of radiation pressure contribution, are shown in left and right column of Figure \ref{fig:LCs_FFT}, respectively. 

Consistent with gas-pressure-only simulations, even with the radiation pressure the emission is brighter in the optical and UV bands, while it is dimmer in the soft X-ray band. Another consistent trend is that variability tends to increase with frequency: while VRO G band flux changes by a factor $\le 1$ order of magnitude, the UV and X-ray fluxes can experience oscillations up to $2$ orders of magnitude. The emission in the optical bands comes mostly from the circumbinary disc, which is colder and less variable than the emission that comes from within the cavity. The high variability found in the UV is due to the high variability of the streams and the mini-discs, which also impacts the emission of the non-thermal hot corona that emits in the soft-X band. The large amplitude variation in the UV flux that we find suggests that these systems might be identifiable by future wide-field UV facilities, such as ULTRASAT \citep[scheduled for launch in 2027,][]{ultrasat} and UVEX  \citep[scheduled for launch in 2030,][]{uvex}.

The gas-pressure-only equal-mass simulations show a clear flux periodicity structure. In the FFT of the fluxes across all bands, we can clearly see the lump periodicity at $0.2\,f_{\rm K}$ and $0.15\,f_{\rm K}$ for the circular $e=0$ and highly eccentric $e=0.9$ case, where $f_{\rm K}$ is the Keplerian frequency of the binary. Moreover, in both cases the FFT peaks associated with the binary motion are present: in the circular case one peak at $f=2\, f_{\rm K}$ and in the eccentric case peaks at $f=1,2\, f_{\rm K}$. 

%%%%%%%%%%%%%%%%%%%%%%%%%%%%% %Figure Surface density and effective temperature maps %%%%%%%%%%%%%%%%%%%%%%%%%%%%%%%%%%%%%%%%%%%%%
\begin{figure}
    \begin{center} 
    \begin{overpic}[width=\columnwidth]{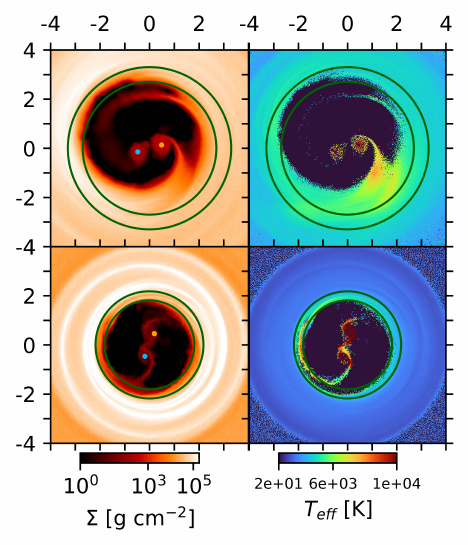}
    \put(35,86){\color{black}\LARGE e=0, q=1}
    \put(-4,68){\rotatebox{90}{\color{black} \LARGE $P_{\rm gas}$ }}
    \put(-4,25){\rotatebox{90}{\color{black} \LARGE $P_{\rm gas} + P_{\rm rad}$ }}
    \end{overpic} \\

    \begin{overpic}[width=\columnwidth]{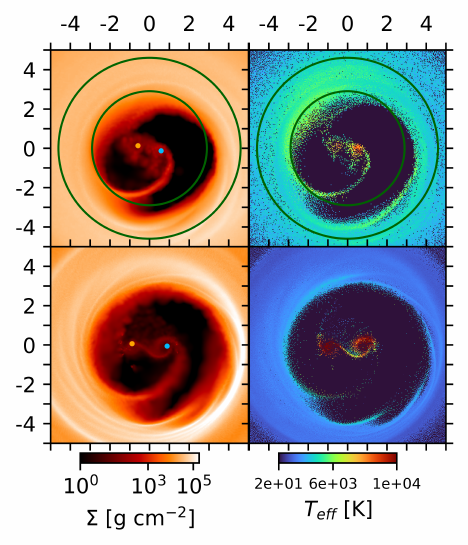}
    \put(31,86){\color{black}\LARGE e=0.9, q=1}
    \put(-4,68){\rotatebox{90}{\color{black} \LARGE $P_{\rm gas}$ }}
    \put(-4,25){\rotatebox{90}{\color{black} \LARGE $P_{\rm gas} + P_{\rm rad}$ }}
    \end{overpic} \\
    
    \caption{Surface density and effective temperature maps for circular equal-mass binaries with $e=0$ (upper figure) and $e=0.9$ ($e=0.9$, $q=1$) binaries (bottom figure). For each case, the first and second row show the result of simulations without and with radiation pressure, respectively. The green circle marks the region from which the main emission modulation in the FFT originates: in gas-pressure-only cases, this corresponds to the lump ($R \sim 2.5 - 3 \,a$ for $e=0$, and $R \sim 2.9 - 4.5 \,a$  for $e=0.9$), while with radiation pressure the main emission at low frequencies in the circular binary arises from the cavity edge ($R \sim 1.81 - 2.18\,a$).
    }
    \label{fig:SigmaT_maps}
    \end{center}
\end{figure}
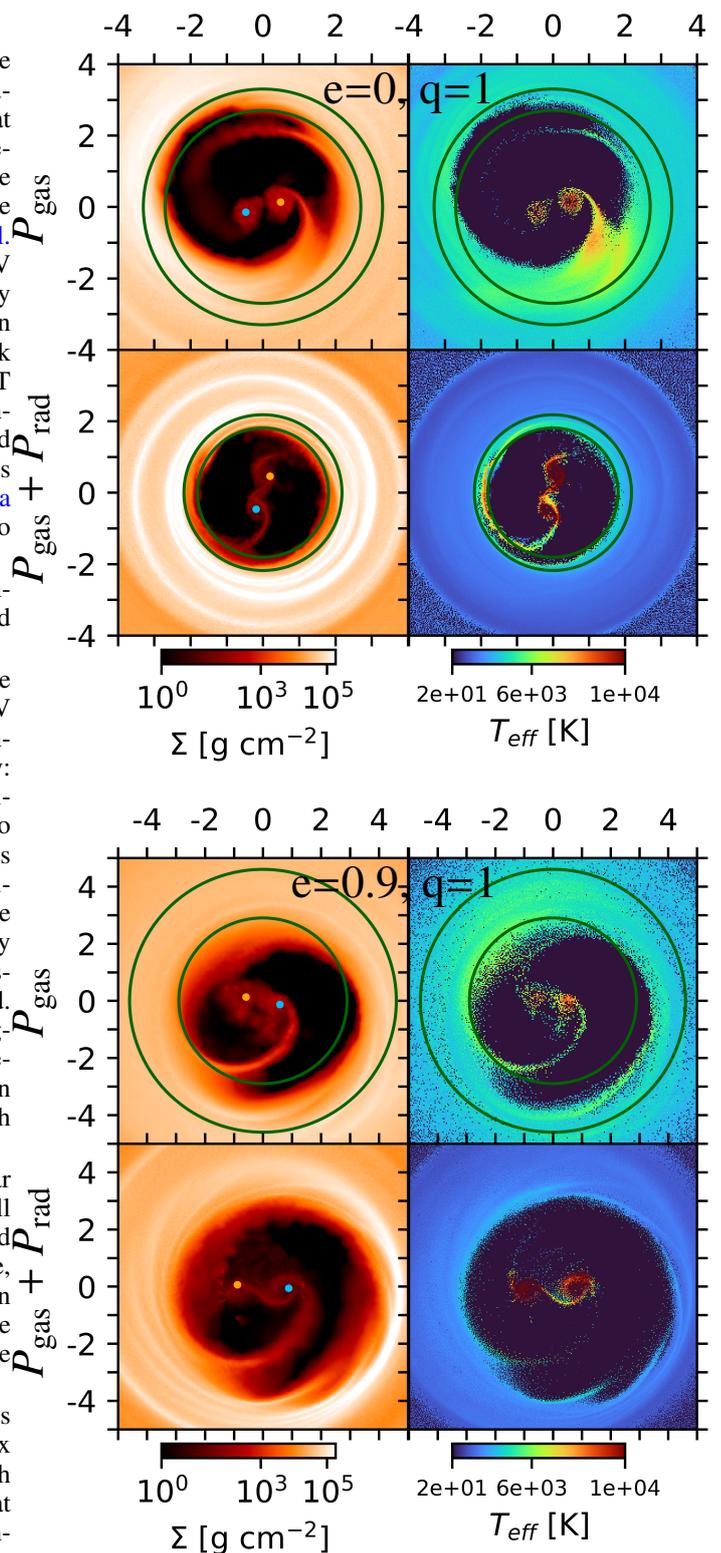

%%%%%%%%%%%%%%%%%%%%%%%%%%%%%%%%%%%%%%%%%%%%%%%%%%%%%%%%%%%%%%%%%%%%%%%%%%% 

When radiation pressure is included, the picture becomes more complex. In the circular equal-mass case, both the flux and the accretion rate FFTs show a more intricate pattern, with a prominent peak at $f = 0.36 \, f_{\rm K}$ with its second harmonic at $ f = 0.72 \, f_{\rm K}$. 
Additionally, since the motion of gas flows is complex and the fluxes are not purely sinusoidal, the interaction between the gas in the mini-discs, streams, and the cavity edge produces additional peaks in the FFT corresponding to linear combination of the two main fundamental modes: the peak at $f = 1.32 f_{\rm K}$ corresponds approximately to the sum of $f = 0.36 \, f_{\rm K}$ and $f = 1.0 \, f_{\rm K}$, while the peak at $f = 1.68 \, f_{\rm K}$ corresponds to a combination of $f = 1 \, f_{\rm K}$ and $f = 2 \times 0.36 f_{\rm K}$. 
In the highly eccentric case $e=0.9$, the FFT displays only the periodicities at $f=f_{\rm K}$ and $f=2f_{\rm K}$ which are driven by the binary orbital motion. Although some power is present near the expected lump frequency, it does not emerge as a distinct peak, unlike in the gas-pressure-only simulations.
Similarly, in the mildly eccentric unequal mass case ($e=0.45$ and $q=0.7$), the flux FFT shows two peaks at $f=f_{\rm K}$ and $f=2f_{\rm K}$, both with and without the radiation pressure contribution. 

In order to better understand the origin of the periodicities in the radiation pressure simulations and the differences with the ones in the gas-pressure-only simulations, we refer to Figure \ref{fig:SigmaT_maps}. The figure displays the surface density and effective temperature maps for circular equal-mass ($e=0$, $q=1$) binaries (upper sub-figure) and highly eccentric equal-mass ($e=0.9$, $q=1$) binaries (bottom sub-figure) at the same time as reported in Figure \ref{fig:SEDs}.
The first and second rows show the surface density map and the effective temperature map for simulation without and with the implementation of the radiation pressure contribution, respectively.

Since the frequency scales with radius as $f \propto R^{-3/2}$, we mark with green circles the regions from which the emission associated to the main peak in the FFT analysis is coming from. In gas-pressure-only simulations, these regions extend over $R \sim 2.5-3 \,a$ for the circular binary, and $R \sim 2.9-4.5 \,a$  for the eccentric case. These correspond to the location of the "lump", with frequency emission peaks at $f=0.2 \, f_{\rm K}$ and $f=0.15 \, f_{\rm K}$, respectively. 

When radiation pressure is included, the FFT emission in the circular case shows a prominent peak at $f=0.36 \, f_{\rm K}$, which originates from a region at $R \sim 1.81-2.18\, a$, i.e., at the edge of the cavity. During the binary accretion, some of the gas that leaks into the cavity is flung back towards the inner edge of the circumbinary disc, where it produces shocks that heat the cavity edge. This gives rise to a circular region with an effective temperature about one order of magnitude higher than that of the circumbinary disc. 
In the highly eccentric case ($e=0.9$), although a slight temperature enhancement is still present at the cavity edge, we do not find any low-frequency modulation and we therefore omit the green circles in the panels of Fig. \ref{fig:SigmaT_maps}. 
It is important to note that, in the circular case, even if the lump modulation does not form, there is still a flux modulation associated with the gas motion along the inner edge of the cavity. However, in the highly eccentric case, the lump is visible in the surface density map, but the temperature remains almost uniform. As a result, no flux modulation is linked to the motion of the cavity edge and there are only the modulations at $f=f_{\rm K}$ and $f=2f_{\rm K}$.
These results suggest that the radiation pressure contribution largely suppresses the presence of the electromagnetic signature associated with the lump-driven flux modulation.

\begin{figure}
    \begin{center}
    \includegraphics[width=1.\columnwidth]{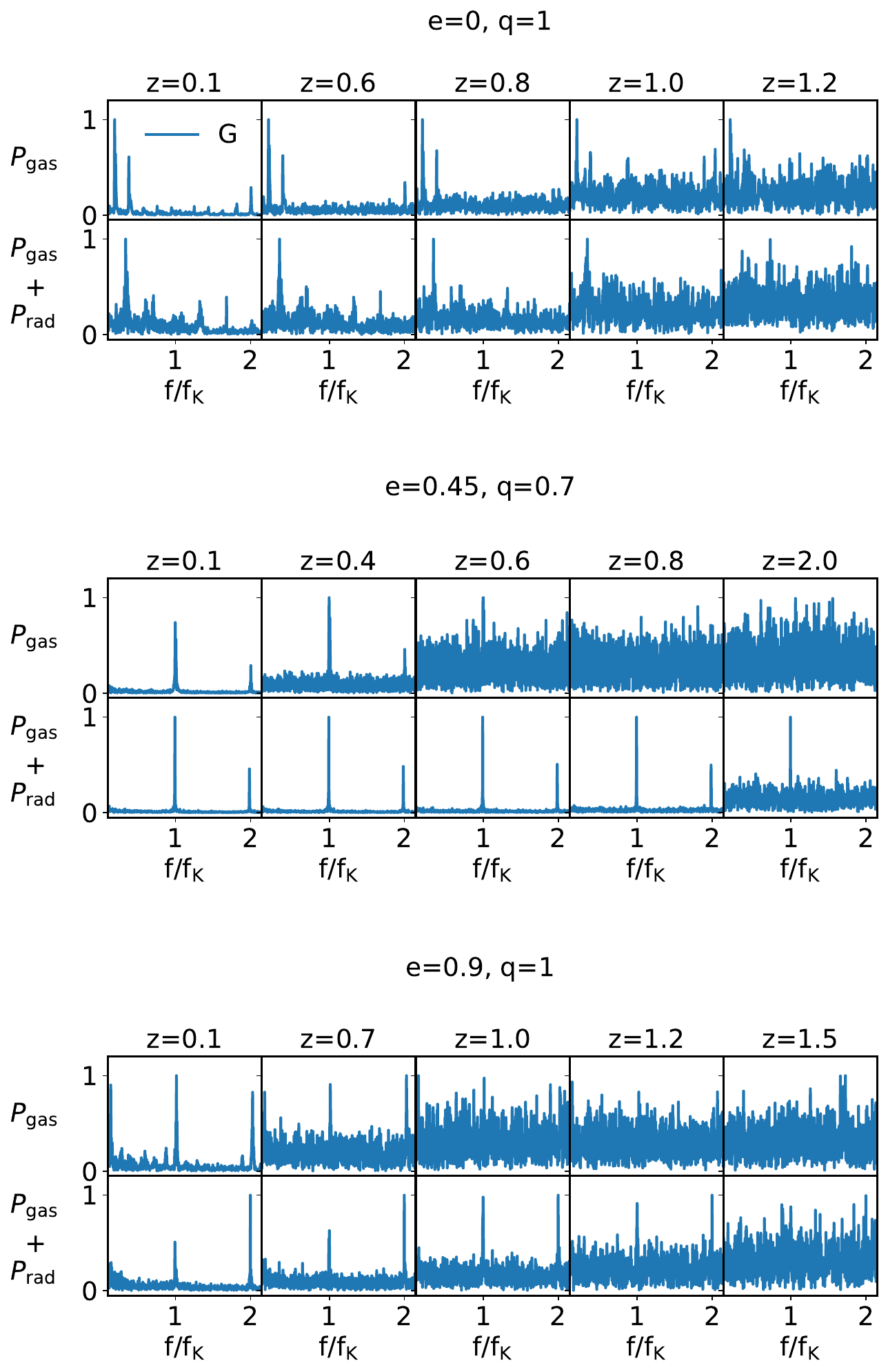}
    \caption{Fast Fourier Transform of the optical G band flux. From the top to the last panel: circular, mildly eccentric $e=0.45$ and highly eccentric $e=0.9$ binaries. The first and second row of each case shows the FFT computed without and with the implementation of the radiation pressure, respectively. The FFT is computed over 300 orbital periods in the eccentric case and 400 orbital periods in the eccentric ones, placing the binaries at different redshifts.}
    \label{fig:FFTs_gband}
    \end{center}
\end{figure}

Finally, we investigate the impact of placing the binary system at different redshifts in order to assess the detection limits taking into account a 5$\sigma$ sensitivity detection threshold in the optical band. 
Figure \ref{fig:FFTs_gband} shows the results. As expected, more distant binaries appear to be fainter and harder to detect. In gas-pressure-only simulations, we find different behaviour based on different cases. In the circular case,  all the periodicities are clearly visible up to redshift $z=0.6$, while at higher redshift only the signal associated with the lump at $f=0.2 \, f_{\rm K}$ remains detectable up to redshift $z=1.2$. In the mildly eccentric case $e=0.45$, all the peaks are distinguishable up to redshift $z=0.4$, whereas beyond this redshift only the modulation associated to the orbital period at frequency $f=f_{\rm K}$ can still be identified up to redshift $z=0.6$. In the highly eccentric case $e=0.9$, the three periodicities can be all detected up to $z=0.7$, but at higher redshift the flux becomes very noisy and the peaks are no longer distinguishable. Instead, the inclusion of radiation pressure improves the detectability, in almost all the cases, in particular in the eccentric ones. %In contrast, with the radiation pressure the results improve: 
In the circular binary case $e=0$, the periodicities remain visible up to $z\sim 0.8$, with the periodicity at $f=0.36 \, f_{\rm K}$ visible up to $z=1.0$, while for highly eccentric system $e=0.9$ they are still detectable up to $z\sim 1.2$. In the mildly eccentric case $e=0.45$, the periodicity associated with the binary orbital period can even be detected up to redshift $z=2$. The enhanced emission from the mini-discs and streams due to radiation pressure, once redshifted, results in a higher flux in the optical bands, improving detectability of MBHBs signatures even at higher redshifts. 

%%%%%%%%%%%%%%%%%%%%%%%%%%%%%%%%%%% FFT gband %%%%%%%%%%%%%%%%%%%%%%%%%%%%%%%%%%%%%%%%%%%%%%%%%%
%%%%%%%%%%%%%%%%%%%%%%%%%%%%%%%%%%%%%%%%%%%%%%%%%%%%%%%%%%%%%%%%%%%%%%%%%%%%%%%%%%%%%

\subsection{Periodic signal identification in VRO survey}

Periodic features in the light curves of MBHBs represent one of the most promising electromagnetic signatures to distinguish them among candidates. Testing whether such signals can be recovered with the cadence and sensitivity of the future VRO is therefore crucial to this aim. So far, we have computed the FFT across a large set of binary orbits, successfully disentangling the different periodicities arising from binaries with different mass rations and eccentricities. However, VRO will likely have access to only a limited number of binary orbits during its 10-years surveys, depending on the binary orbital period. Thus, it could be non-trivial to detect periodicities. 

\begin{figure}
    \begin{center}
    \includegraphics[width=\columnwidth]{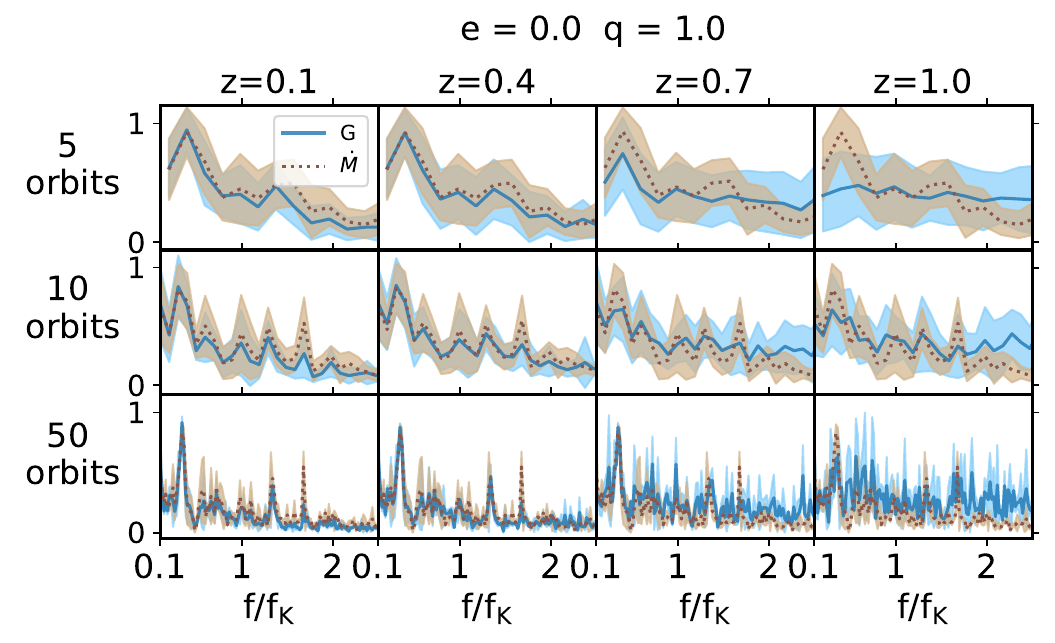} \\
    \includegraphics[width=\columnwidth]{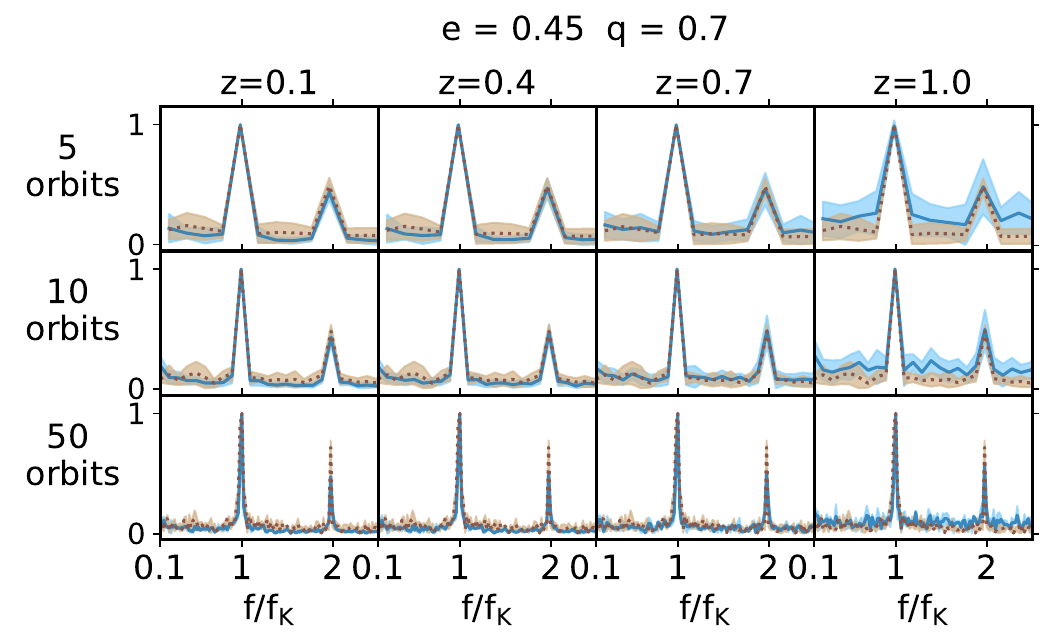}
    \\
    \includegraphics[width=\columnwidth]{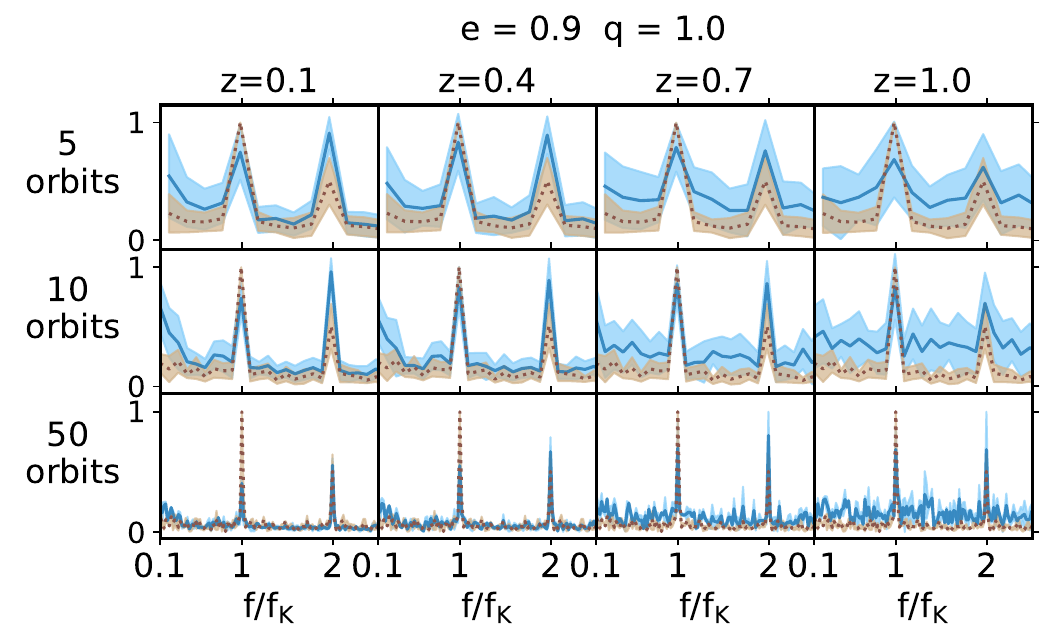}
    \caption{Fast Fourier transform of LCs from simulations with the implementation of radiation pressure. From the top to the last panel: circular, mildly eccentric $e = 0.45$ and highly eccentric $e = 0.9$ binaries. The first row of each case shows the FFT of the Optical G band flux (blue line) and the FFT of the accretion rate (brown line) computed over 5-orbit windows within a total of 300/400 orbital periods at redshift z = 0.1, 0.4, 0.7, 1.0. The second and the third rows show the FFT computed over 10- and 50-orbit windows, respectively.}
    \label{fig:FFTs_mean}
    \end{center}
\end{figure}

In order to evaluate the impact of the radiation pressure on the periodicity detectability by the VRO, we repeated the same analysis carried out in \cite{Cocchiararo2024}: we perform FFT of the accretion rate and flux in the optical G band over 5,10 and 50 orbits windows, sliding them across a total of 300 and 400 orbits in circular and eccentric binaries, respectively. For each configuration, we average the FFT and compute their mean and standard deviation at redshifts $z=0.1, 0.4, 0.7,1.0,1.5,2$. The results are shown in Figure \ref{fig:FFTs_mean} where, for simplicity, we only show %are shown only 
the cases with redshift $z=0.1,0.4,0.7,1.0$. Comparing our results with those in Figure 6 from \cite{Cocchiararo2024}, we find the same main trend observed in gas-pressure-only simulations: periodicities are easier to detect in unequal-mass systems, while equal-mass binaries exhibit weaker signals that improve with longer time-windows. As in the gas-pressure-only case, at higher redshifts, the variance, i.e. the statistical error on the mean flux/accretion rate, grows due to the VRO sensitivity limit, particularly in the equal mass and highly eccentric binaries.

In the equal-mass case, only the main peak at $f \sim 0.36 \, f_{\rm k}$ can be distinguished at low redshift ($z<0.4$) when considering 5-10 orbits, while increasing the number of orbits up to 50 makes the identification of periodicities possible but only up to $z\sim 0.7$. %This binary case represent the most challenging configuration to detect at higher redshift among our models. 
In the highly eccentric equal-mass case ($e=0.9$, $q=1$) %the radiation pressure suppresses the lump flux variability and we observe only two main periodicities, at $f=1,2\, f_{\rm K}$. We observe a similar result that we can find in the unequal-mass binaries: 
the peaks corresponding to the two modulations at $f=f_{\rm K}$ and $f=2f_{\rm K}$ are more pronounced and can be clearly identified even taking into account a few orbits, in contrast to the case of gas-pressure-only simulation, where the signals are distinguishable only taking into account at least 10 orbits or 50 orbits at redshift $z<0.4$ and $z<0.7$, respectively. 

In the unequal-mass case, the periodicities are easily distinguishable even at $z=1.5$, even taking into account 5 orbital periods, in contrast to what was found in the gas-pressure-only simulations. 
We find that periodic signatures remain visible up to redshift $z \gtrsim 0.6-1.0$, higher than in the gas-pressure-only case. This improvement is given by the enhanced emission from the mini-discs and streams, which once redshifted boosts the observed flux in the optical band. Neglecting radiation pressure in the model may artificially lower the predicted detectability rate.

We note that we have assumed 5 and 10 orbits windows for the FFT because our binaries have a $\sim 1$ year orbital period and VRO would last 10 years. 
This choice reflects a compromise: several cycles are needed to detect reliable periodicity, but binaries with longer periods are more numerous and complete fewer cycles within the survey baseline, making them harder to identify. Conversely, binaries with separations sufficiently small to do many cycles in the VRO time-span can instead be more easily detected. 
Finally, a comprehensive statistical evaluation of the VRO ability to detect these periodicity will be carried out in a future work F. Cocchiararo et al. (in prep).

%%%%%%%%%%%%%%%%%%%%%%%%%%%%%%%%%%% LCS WINDOWS %%%%%%%%%%%%%%%%%%%%%%%%%%%%%%%%%%%%%%%%%%%%%%%%%%
%%%%%%%%%%%%%%%%%%%%%%%%%%%%%%%%%%%%%%%%%%%%%%%%%%%%%%%%%%%%%%%%%%%%%%%%%%%%%%%%%%%%%

%%%%%%%%%%%%%%%%% CONCLUSIONS %%%%%%%%%%%%%%%%%%

\section{Conclusions}
\label{sec:conclusions}

In this work, we investigated how radiation pressure affects the electromagnetic signature of milli-parsec-scale MBHBs surrounded by a thin gaseous circumbinary disc using 3D numerical simulations with hyper-Lagrangian refinement. We modelled the disc with an adiabatic equation of state, allowing the gas temperature to evolve through shocks, PdV work, and black body cooling. We considered systems with eccentricity $e=0, 0.45, 0.9$ and mass ratio $q=1, 0.7$. We investigated the role of radiation pressure in shaping the SEDs and the multiwavelength LCs, focusing on the detectability of periodic signature in surveys like the future VRO. 

We find that radiation pressure plays a crucial role in shaping both the spectral energy distributions and the temporal variability of MBHBs, affecting the electromagnetic signatures of these systems. 
%We computed the SEDs from the circumbinary disc, assuming a black-body emission. 
Radiation pressure produces opposite effects on the different emitting regions of the system. While the circumbinary disc becomes geometrically thinner, colder and less luminous with the emission peak shifting towards lower frequencies ($\Delta \log\, (\nu/\rm Hz) \sim 0.2-0.3$), the mini-discs experience vertical expansion and have a higher effective temperature, resulting in a shifted emission toward higher frequencies ($\Delta \log\, (\nu/\rm Hz) \sim 0.2-0.3$ in circular binaries and $\Delta \log\, (\nu/\rm Hz) \sim 0.4-0.6$ in eccentric cases) with luminosities increased by $1-2$ orders of magnitude with respect to the gas-pressure-only simulations. %In the simulations with only gas pressure, the mini-discs emission is completely covered by the streams emission but when radiation pressure is included, the mini-disc emission is no longer covered at least in the moderate eccentric case $e=0.45$, highlighting the enchanted contributions of the mini discs to the total luminosity. 

%The impact of radiation pressure on the spectral emission is mirrored as well in the modulation of the flux of these systems. 
The inclusion of the radiation pressure contribution modifies also the modulations of the flux coming from these systems.
We computed the LCs in different frequency bands, mainly focusing on the optical window that will be probed by VRO. % We calculated the thermal flux emitted in the optical frequency bands using the VRO filters %\af{if the following is already in the results section remove it here, otherwise move it to the results section}, in the near-UV band within $1.0-1.5 \, \times 10^{15} \, \rm Hz$ $(4.13\,-6.20 \, \rm eV)$, and in the soft-X band in the range of $7.25-48.3 \, \times 10^{16} \, \rm Hz$ $(0.3-2 \, \rm keV).$. 
Consistently with the shift in emission frequency, we find that the flux in the optical and near-UV frequency bands is higher with respect to the gas-pressure only simulation cases. 
The variability is also affected: while in the optical G band, flux oscillations reach a factor $\le 1$ order of magnitude, in the near-UV and soft-X ray bands they can vary by up to $2$ order of magnitude as in gas-only simulations.

In contrast to gas-pressure-only simulations  \citep{Cocchiararo2024}, where a clear modulation associated to the lump was found in %both the equal-mass binaries (with frequency peak at $f=0.2, 0.15$ in the 
equal-mass circular and highly eccentric binaries, here the lump-driven modulation is strongly suppressed. 
In the circular equal-mass case, both the flux and the accretion rate FFTs reveal clear peaks at  $f\sim 0.36 \, f_{\rm K}$ and $f\sim 0.72 \, f_{\rm K}$, associated to emission from the heated inner edge of the cavity. 
Indeed, during the accretion of gas onto the binary, some gas is flung back to the cavity edge, generating shocks that locally heat the gas. Moreover, in the LCs, we can notice other peaks: at $f=f_{\rm K}$ and $f=2f_{\rm K}$ corresponding to a periodicity of the binary orbital period and half of it and $f=1.32, 1.68 \, f_{\rm K}$ interpreted as linear combinations of the main two fundamental modes. %\af{see comment above on beat frequency}
In the eccentric cases, the FFT shows only the periodicities at  $f=f_{\rm K}$ and $f=2f_{\rm K}$.

The results discussed so far apply when the FFT is evaluated over long time spans, corresponding to several hundreds of binary orbits. However, VRO will cover only a decade of observations, while most compact MBHBs are expected to have orbital periods of the order of years. To provide a more realistic estimate of the detectability of periodic signals, we therefore computed the FFT of the flux and accretion rate using shorter time windows of 5, 10, and 50 orbits, putting the source at different redshifts, and we compared the results with those obtained from gas-pressure-only simulations. Even with radiation pressure, in the circular equal-mass case identifying periodic signals is non-trivial, in particular with a handful of orbits. As expected, the periodic peaks become clearer and the statistical noise decreases when the number of orbits increases. In the highly eccentric equal-mass case the radiation pressure suppresses the lump modulation, but variability on the binary orbital period is quite pronounced, and the detection of periodicities even with a few orbits is improved with respect to the gas-pressure-only case. Finally, in the unequal-mass case, as also reported in the gas-pressure-only simulations, periodicities are easier to detect even at higher redshift $z\sim 1.5$.
Because of the relatively low luminosity of our systems ($L_{\rm bol} \approx 10^{42} \, \rm erg \, s^{-1}$), periodic signals would be observable with VRO only from systems at redshift $z<0.7-1.0$.

The suppression of the lump modulation in equal-mass systems once radiation pressure is included in hydrodynamical simulations challenges one of the most promising EM tracers suggested by previous works. This results stress the importance of adopting simulations with more realistic physical processes in order to obtain a more comprehensive model of these systems and better characterise their EM signatures. At the same time, it is crucial to developed new statistical approaches to assess the observability of the predicted features. This second point will be the subject of a future publication %\al{see above, if it is a future publication, it cannot have a citation yet}
Cocchiararo et al. (in prep).

%%%%%%%%%%%%%%%%% ACNOWLEDGEMENTS %%%%%%%%%%%%%%%%%%
\section*{Acknowledgements}

We thank Daniel Price for providing the {\sc phantom} code for numerical simulations and acknowledge the use of {\sc splash} \citep{Price2007} for the rendering of the figures.
We thank Phil Hopkins for providing the {\sc gizmo} code for numerical simulations. 
AS acknowledges financial support provided under the European Union’s H2020 ERC Consolidator Grant ``Binary Massive Black Hole Astrophysics" (B Massive, Grant Agreement: 818691). AS also acknowledges the financial support provided under the European Union’s H2020 ERC Advanced Grant ``PINGU'' (Grant Agreement: 101142079). AL acknowledges support by the PRIN MUR "2022935STW".
AF acknowledges financial support from the Unione europea- Next Generation EU, Missione 4 Componente 1
CUP G43C24002290001. 

%%%%%%%%%%%%%%%%%%%%%%%%%%%%%%%%%%%%%%%%%%%%%%%%%%

\bibliographystyle{aa} % style aa.bst
\bibliography{bibliography}
\end{document}